\documentclass[12pt,dvips]{article}
\usepackage{amssymb}
\usepackage{amsmath}
\usepackage{epsfig}
\begin{document}

\title{Structure-generating mechanisms in agent-based models}
\author{R. Vilela Mendes\thanks{%
Grupo de F\'{\i }sica Matem\'{a}tica, Complexo Interdisciplinar,
Universidade de Lisboa, Av. Gama Pinto 2, 1699 Lisboa Codex, Portugal} 
\thanks{%
Zentrum f\"{u}r interdisziplin\"{a}re Forschung, Universit\"{a}t Bielefeld,
Wellenberg 1, 33615 Bielefeld, Deutschland} \thanks{%
e-mail: vilela@alf1.cii.fc.ul.pt}}
\date{}
\maketitle

\begin{abstract}
The emergence of dynamical structures in multi-agent systems is analysed.
Three different mechanisms are identified, namely: (1) sensitive-dependence
and convex coupling, (2) sensitive-dependence and extremal dynamics and (3)
interaction through a collectively generated field. The dynamical origin of
the emergent structures is traced back either to a modification, by
interaction, of the Lyapunov spectrum or to multistable dynamics.
\end{abstract}


\section{Introduction}

Organization and structure are ubiquitous in natural phenomena and much of
the scientific endeavor is aimed at the discovery of patterns in the raw
data supplied by Nature. Whenever a pattern is detected, it can be used to
obtain a compressed description of the phenomenon or to predict its outcome.
Prediction through compressed descriptions is also the way many living
beings deal with the external world, ants included\cite{Ants}. If a natural
process is stationary, the discovery of its patterns by an observer is of no
consequence for the system itself. It is merely an information processing
feature of the observer. However, for evolving composite systems, the
emergence of collective patterns may be a determining factor in the process
of coevolution.

To understand what a pattern (or structure) is, how it is represented and
how it may be used for prediction, has been the subject of a great deal of
research in the past. This ranges from time series prediction (for reviews
see \cite{Santana} \cite{Weigend}), to stochastic model identification\cite
{Ornstein} \cite{Kazakos} \cite{Amari1}, to dynamical system reconstruction%
\cite{Farmer1} \cite{Crutch1} , to coding\cite{Ziv} \cite{Rissanen} and to
the quantitative characterization of the complexity of patterns\cite
{Grassberger} \cite{Crutch2} \cite{Vilela4}. Of course, references here
cannot be exhaustive nor make justice to a very large body of interesting
work. They are only representative of the type of work developed in each
approach.

Concerning the question of what a structure is and how it may be used for
prediction, a good part of the work done in the past fits in the general
scheme of Crutchfield's \textit{computational mechanics}\cite{Crutch3}. A
different question, however, concerns the dynamical mechanisms by which
collective structures arise on composite systems. And also why their
dynamical behavior may be so different from the dynamics of the components
when in isolation.

Cross and Hohenberg\cite{Cross} discussed pattern formation in systems
modelled by partial differential equations, by analyzing the instabilities
of the homogeneous states. Near the instabilities the dynamics is described
by amplitude equations which characterize the collective variables. Here a
different approach is followed. Systems composed of many agents in
interaction, each one having \textit{simple dynamics}, are considered. No
collective variables are built in the model to begin with, and any
collective structures that may be observed should appear as emergent
properties of the dynamical process.

What is meant by simple dynamics of the agents must however be clarified.
The idea is that the dynamics is simple to describe in law, but not that it
has simple orbits. In short, a dynamical law with small \textit{%
sophistication}\cite{Koppel}, but capable of generating orbits of high
Kolmogorov complexity. A paradigmatic example is multiplication by $p$ (mod.$%
1$) $\left( p=2,3,...\right) $. 
\begin{equation}
x_{n+1}=px_{n}\quad (\textnormal{mod}.1)  \label{1.1}
\end{equation}
It has an invariant measure absolutely continuous with respect to Lebesgue,
positive Lyapunov exponents and Kolmogorov entropy, as well as orbits of all
types. It is clear that, even if simple to describe, the dynamics of each
agent must contain enough dynamical freedom for interesting behavior to be
obtained when the agents are put in interaction. Otherwise, if dead rocks
are added to a barren landscape, all one obtains is a rocky ground.

In a multi-agent system, a temporal or spatial structure is defined as a
phenomenon which has a time or space scale much larger than the
corresponding scales of the individual agent dynamics. Structures and other
collective properties of multi-agent systems may be rigorously characterized
by ergodic invariants\cite{Vilela1}-\cite{Vilela3}. A short summary of the
main ergodic invariants that may be used to characterize multi-agent systems
is included in the Appendix. When a unique invariant measure controls the
dynamics, the ergodic invariants provide an adequate characterization of the
system behavior. However in multistable systems, many different measures
must be taken into account and a different type of parameters must be used.
This is also sketched in the Appendix.

In the same way as there is no unique way to characterize the complexity of
dynamical systems, each feature requiring a different complexity parameter%
\cite{Grassberger} \cite{Vilela4}, one should not expect to find a unique
universal mechanism responsible for all structure-generating effects. In
this paper three different mechanisms are identified, of which some examples
are studied, namely

(1) Sensitive-dependence and convex coupling

(2) Sensitive-dependence and extremal dynamics

(3) Interaction through a collectively generated field (Multistability and
evolution)

In some of these mechanisms an important role is played by
sensitive-dependence, that is, by the fact that the individual agent
dynamics has positive Lyapunov exponents, as in Eq.(\ref{1.1}). Without
interaction the system would have a degenerate positive Lyapunov spectrum.
The interaction lifts the degeneracy and it is the fact that different
directions in phase-space acquire different separation dynamics that creates
the collective dynamical structures. In particular when a Lyapunov exponent
approaches zero from above, it creates a feature with a very long time
scale. The modification of the Lyapunov spectrum arises either from varying
interaction strength or from extremal dynamics, that is, the mechanism by
which only the agent under the largest stress is allowed to evolve.

A remarkable exception to the above paradigm of Lyapunov spectrum
modification occurs when there are no direct interactions between the
agents, which only react to a collective variable, that they themselves
create. In this case, sensitive-dependence of the agent dynamics influences
the fluctuations, but self-organization and collective variables are mainly
controlled by evolution processes and multistability of the dynamics.

\section{Sensitive-dependence and convex coupling}

Here the individual agent dynamics is assumed to have positive Lyapunov
exponents. Without interaction, the system would have a degenerate spectrum
of positive Lyapunov exponents. Convex coupling, as in Eq.(\ref{2.1}), has a
contractive effect. Therefore, for sufficiently large interaction strength,
some of the Lyapunov exponents approach zero from above. Physically, the
mechanism of convex coupling relates, for example, to a situation where
there is a limitation on the range of options and influences that determine
the actions of the agents. Then if an agent receives an influence from
someone else there is a correspondent decrease of the influence of its own
state in the future evolution. On the other hand the intensity of the
interactions between the agents may depend on the history of past
interactions or on the number of agents that occupy the same volume of space.

This mechanism will now be illustrated by one example. It deals with a
situation where the number of interacting agents varies, according to a
reproduction and death scheme, and the strength of interaction depends on
the number of agents at any given time. Of particular interest is the
population control effect of the correlations.

The model is a system of Bernoulli agents on a circle with nearest-neighbor
interactions by convex coupling, namely

\begin{equation}
x_{i}(t+1)=(1-c)f(x_{i}(t))+\frac{c}{2}\left(
f(x_{i+1}(t))+f(x_{i-1}(t))\right)  \label{2.1}
\end{equation}
with $f(x)=2x$\ (mod. $1$) and periodic boundary conditions. The agents are
assumed to live in a limited space, the intensity of the coupling being a
function of the total number $N$ of agents, for example 
\begin{equation}
c=c_{m}\left( 1-e^{-\alpha N}\right)  \label{2.2}
\end{equation}

With fixed coupling, this model was used in the past to illustrate the
behavior of the ergodic invariants for self-organization\cite{Vilela3} and
also in several other studies of the dynamics of coupled map lattices.

Here the coupling becomes a dynamical variable as well, by a \textit{%
reproduction and death} mechanism defined as follows:

- After each $R$ time cycles, the system is examined, agents which at that
moment have $x_{i}>0.5$ are coded $1$ and those for which $x_{i}\leq 0.5$
are coded $0$.

- Then, configurations $0110$ are candidates for reproduction with
probability $p_{r}$ and configurations $0000$ are candidates for death with
probability $p_{m}$.

- Reproduction is the transition $0110$ $\rightarrow $ $0X110$ with the
state of the new agent $X$ being chosen at random in the interval $\left(
0,1\right) $.

- Death is the transition $0000$ $\rightarrow $ $000$.

It is clear that without coupling the two configurations $0110$ and $0000$
appear, on average, the same number of times and the variation of the
population density depends only on the relative values of $p_{r}$ and $p_{m}$%
. With coupling the situation will be different and the model shows how
correlations, generated by coupling, influence the inter-agent evolution
mechanism.

In this example a rigorous characterization is possible of the structures
that develop through interaction. As explained in \cite{Vilela3} this
characterization is obtained through the computation of the Lyapunov
exponents from which a \textit{structure index} is constructed (see the
Appendix). Further insight is obtained from conditional exponents as well,
but they will not be used here. The Lyapunov exponents for the dynamical
system in (\ref{2.1}) are 
\begin{equation}
\lambda _{k}=\log \left\{ 2\left( 1-c\right) +2c\cos \left( \frac{2\pi }{n}%
k\right) \right\}  \label{2.3}
\end{equation}
$k=0,\cdots ,N-1$. They are all positive for $c<0.5$ and when the coupling
varies above this value one observes the crossing through zero of each
individual Lyapunov exponent and successive changes in the structure of the
system. That means that, for $c\neq 0$, each collective mode has a different
probability, a collective mode being frozen each time a Lyapunov exponent
reaches the zero value.

The eigenvectors corresponding to each exponent are $\left\{ e^{in\theta
_{k}}\right\} $ with $\theta _{k}=\frac{2\pi }{N}k$ , $k=0,\cdots ,N-1$.
Therefore 
\begin{equation}
y_{k}=\frac{1}{N}\sum_{n=1}^{N}\cos \left( \frac{2\pi }{N}kn\right)
\label{2.4}
\end{equation}
are the coordinates of the collective eigenmodes. In Fig.1 one shows the
average energy $E_{k}=\left\langle y_{k}^{2}\right\rangle $ of the
collective modes for several values of the coupling (for a system with $100$
agents). In all plots the mode $k=0$, that is just the average over all
agents, is not shown. The expected suppression and freezing of many modes is
quite apparent. Notice that some modes that are suppressed for certain
values of the coupling become restored for higher values. This is apparent,
for example, for $c=1$.


Fig.2 shows the evolution of the population plotted against the
reproduction-death cycle number. The probabilities are $p_{r}=1$ and $%
p_{m}=0.5$. This is a situation for which, without coupling, the population
would grow indefinitely. However with the density-dependent coupling (\ref
{2.2}) the population becomes controlled with fluctuations around some
average value. Three of the plots show this stabilization starting from
different initial conditions. Even if the population stabilizing dynamical
mechanism leads to a non-zero average value, a large fluctuation may lead to
extinction, as shown in the last plot.


The population stabilizing mechanism is a consequence of the correlations
that develop as a result of the unequal distribution of energy among the
collective modes, caused by the coupling. Fig.3 shows the relative
probability of each one of the $16$ different configurations of four
neighbors ($x_{1}x_{2}x_{3}x_{4}$), labelled by $x_{1}+2\times x_{2}+4\times
x_{2}+8\times x_{2}$. As seen in the first plot, without coupling all
configurations are equiprobable.


Changes in the dynamical structure appear associated to the points where
each Lyapunov exponent crosses zero. These are the points where the
structure index diverges because the time scale associated to the vanishing
Lyapunov exponent becomes infinite. In this sense these points are similar
to statistical mechanics transition points. This transition is even more
dramatic when a large number of Lyapunov exponents crosses zero
simultaneously. This is, for example, the case for the globally coupled
model studied in Ref.\cite{Vilela1}. The transition points correspond to
well defined (sharp) values of the parameters (coupling parameters,
population density, etc.). A system may approach such points in the course
of its life, by dynamical evolution of the population density, as seen
above, or by environment changes. However, to generate a spontaneous
approach to criticality, without fine tuning of the parameters, a different
mechanism is required. As seen later on, extremal dynamics together with
positive Lyapunov exponents of the individual dynamics is such a mechanism
that drives a system to the edge of criticality, that is, to the point where
the Lyapunov exponents approach zero.

\section{Criticality and extremal dynamics}

Here one analyses the case where the individual dynamics is sensitive
dependent, that is, it has at least one positive Lyapunov exponent, but the
collective dynamics is of the extremal type. That means that at each time
step only the agent under the largest stress is allowed to evolve. Depending
on the specification of the dynamics the largest stress may be the largest
driving force or the largest or smallest value of a state variable. For
example in the Bak-Sneppen\cite{Bak1} model the agent that evolves (together
with its neighbors) is the one that has the smallest value of the state
variable and in the train model\cite{Vieira1} \cite{Vieira2} the agent that
moves is the one that suffers the greatest driving force.

The prescription of extremal dynamics is a feature that simulates friction
or resistance to change in the dynamical system. When parallel dynamics is
replaced by these type of extremal dynamics, a dramatic effect takes place
in the spectrum of Lyapunov exponents, in the limit of a large number $N$ of
agents. For the computation of the Lyapunov exponents of the coupled system,
instead of a tangent map matrix involving all the partial derivatives, one
has now the product of matrices which have ones on the diagonal almost
everywhere and only one non-trivial $r\times r$ block, $r$ being the number
of neighbors that evolve at each time step. Therefore the Lyapunov exponents
are obtained, on average by the root $\frac{r}{N}$ of the $r\times r$
blocks. Therefore if the exponents of the $r\times r$ blocks are positive
then, for large $N$, all the Lyapunov exponents approach zero from above,
independently of any other characteristics of the dynamics. Hence
sensitive-dependence of the individual dynamics plus extremal dynamics leads
the system to the edge of criticality, in the sense that vanishing Lyapunov
exponents means that there is no natural time scale for the separation
dynamics. Recall that these are the points where the \textit{structure index}
diverges (see the Appendix).

Models of these type belong to the general class of \textit{self-organized
criticality} (SOC)\cite{Bak2} although not all SOC models that have been
proposed display the above described mechanism in all its purity. At the end
of the section a comment will be made about this.

A great deal of work, on the SOC phenomenon, has been done by many authors.
Here I only want to emphasize those features that relate to the Lyapunov
spectrum. For definiteness I will concentrate on a continuous version of the
Bak-Sneppen model\cite{Vilela3} which is a $C^{\infty }-$dynamical system
defined as follows:

Let $\overrightarrow{x}\in [0,1)^{N}$ be the vector of coordinates of the
agents and $\Gamma _{i}(\overrightarrow{x})$ the function 
\begin{equation}
\Gamma _{i}(\overrightarrow{x})=\prod_{j=i-n_{V}}^{j=i+n_{V}}\prod_{k\neq
j}\left( 1+e^{\alpha (x_{k}-x_{j})}\right) ^{-1}  \label{3.1}
\end{equation}
For large $\alpha $ the function is nearly zero if $i$ is in the $n_{V}-$%
neighborhood of the agent with the minimum coordinate $(x_{\min })$ and is
nearly one otherwise. The dynamics of the model is defined by 
\begin{equation}
x_{i}(t+1)=\Gamma _{i}(\overrightarrow{x})x_{i}(t)+\left( 1-\Gamma _{i}(%
\overrightarrow{x})\right) f\left( x_{i}(t)\right)  \label{3.2}
\end{equation}
with $f(x)=2x$\ (mod. $1$). For $n_{V}=1$ and in the limit of large $\alpha $
this is equivalent to the original Bak-Sneppen model\cite{Bak1}. Notice that
the usual operation of finding the agent with the smallest barrier is
replaced here by an infinitely differentiable operation and all the
dynamical system techniques and results may be safely applied.

For large $\alpha $ the Lyapunov exponents are 
\begin{equation}
\begin{array}{llllll}
\lambda & \eqsim & \log \left( 2\right) ^{\frac{3}{N}} &  & N & 
\textnormal{times}
\end{array}
\label{3.3}
\end{equation}
Therefore, as stated before, as $N\rightarrow \infty $, $\lambda \rightarrow
0^{+}$. Hence, in this limit, there being no natural scale for the dynamics,
exponential decay terms must disappear in all relaxation phenomena, leaving
only the power law pre-factors.

This dynamical system displays some unusual features, related in particular
to the nature of what has been improperly called its ''attractor''. In the $%
N\rightarrow \infty $ limit the one-agent probability density is uniform
above 0.67 and zero below this threshold. The marginal density for any
finite number $n$ of agents is the projection on a $n-$dimensional
hyperplane of the $N-$dimensional hypercube of side (1-0.67). This hypercube
however is not an attractor because it carries zero or negligible (for
finite $N$) measure. Also it is not a repeller because there are many
neutral directions corresponding to the directions that are not neighbors to
the minimum coordinate nor even a weak repeller because it is not an
invariant set. I will call such a set a \textit{ghost weak repeller}, sets
of this type being characterized by the following conditions:

(i) existence of repelling and neutral directions

(ii) zero or negligible measure

(iii) full measure on the projection to hyperplanes up to dimension $N-k$, $%
N $ being the dimensionality of the system and $k$ a finite number.

Fig.4 illustrates some of these properties for the model defined in Eq.(\ref
{3.2}) with $N=100$, $n_{V}=1$ and $\alpha =1000$.\ The figure shows the
one- and two-agent marginal distributions, the distribution of the distances
to the hypercube of side (1-0.67) and finally the scaling of the avalanches.
The third plot emphasizes the negligible measure that is carried by the
hypercube itself, to be compared with the structure of the marginal
distributions.


Whereas in the model discussed in the previous section, the creation of new
structures, associated to the modification of the Lyapunov spectrum,
occurred at particular values of the parameters, here no fine tuning of the
parameters is needed, the system self-organizing itself spontaneously in a
critical state. In the BS-model, as stated above, the establishment of the
long-range temporal correlations is very naturally associated to the
disappearance in the $N\rightarrow \infty $ limit of any characteristic time
scale in the Lyapunov spectrum. Therefore the BS model is a very
paradigmatic example of the SOC phenomenon in the sense that it always
become critical in the $N\rightarrow \infty $ limit, independently of any
other details (value of $n_{V}$, function $f$ chosen for the dynamics of $%
x_{\min }$, etc.). This may not be the case for all SOC models that have
been proposed.

For example, Zhang's model\cite{Zhang1} was studied as a dynamical system by
Blanchard, Cessac and Kr\"{u}ger\cite{Blanchard}. In their formalism,
Zhang's model is a dynamical system of skew-product type with the first
factor corresponding to the activation of one site and the second to the
energy relaxation process. Their unit of time is not the natural unit of
time, instead each evolution step corresponds to an activation and a full
avalanche. Therefore each step corresponds to different lengths of physical
time, depending on the size of the avalanche. In their step units, there is
one positive Lyapunov exponent ($\log N$) corresponding to the activation
dynamics, the rest of the dynamics being contractive. Denote by $\overline{t}
$ the average number of iteration steps (average duration of an avalanche),
by $\overline{a}$ the average number of distinct relaxing sites in one
avalanche and by $d$ the dimension of the lattice. Then, one estimates that
the expanding Lyapunov exponent in physical time units is of order$\frac{1}{%
\overline{t}}\log \left( N\right) $. If $\lambda _{i}$ is one of the
contracting Lyapunov exponents, in physical units it is of order $\frac{%
\overline{a}(2d+1)}{N}\lambda _{i}$. Therefore all characteristic time
scales will disappear only if $\overline{t}$ grows faster than $\log \left(
N\right) $ when $N\rightarrow \infty $ and $\overline{a}$ grows slower than $%
N$ when $N\rightarrow \infty $. This means that in this model there may be
ranges of parameters for which the SOC phenomenon is not observed. A similar
conclusion is reached by the authors\cite{Blanchard} using other approach.

In the discrete version of the train model proposed by Vieira\cite{Vieira2}
the activation dynamics, being the addition of a fixed quantity $\delta f$
to the first block, is neutral as far as the Jacobian is concerned. The only
non-zero contribution to the Lyapunov comes from the derivative of the
updating function $\phi ^{\prime }$ . It has been noticed by the author\cite
{Vieira2} that SOC behavior is only obtained if $\left| \phi ^{\prime
}\right| >1$. With the estimate $\frac{1}{N}\log \left| \phi ^{\prime
}\right| $ for the global Lyapunov exponent, in the $N\rightarrow \infty $
limit, this case is seen to correspond, once again, to an approach to zero
from above.

\section{Interaction through collective variables. Multistability and
evolution}

In all models studied before, there is some sort of direct interaction
between the agents giving rise to the collective behavior. Another class of
models is the one where the interaction is mediated by a collective
variable. On the other hand, the collective variable is an aggregate result
of the state variables and actions of the agents. Therefore the agents react
to an aggregate variable that they themselves create.

In most models of this type, in addition to the interaction through the
collective variable, there is also an evolution mechanism which plays an
important role in organizing the system. Therefore the dynamics of the
system is a composition of two dynamical laws. One is the (fast) dynamics of
interaction through the collective variable, the other the (slow) evolution
dynamics. Sometimes it turns out that the essential mechanism
self-organizing the system is the evolution mechanism, a slow dynamics,
whereas the fast dynamics only provides the multi-attractor background which
is selected by the slow evolution. As an example of this mechanism two
models will be studied.

\subsection{Coupled map minority model}

Inspired on Brian Arthur's bar model\cite{Arthur1}, models have been proposed%
\cite{Zhang2} \cite{Marsili} \cite{Cavagna1} where the agents choose the
value of a variable ($\pm 1$, for example) and those that are on the
minority group win a point. In a continuous version\cite{Vilela3}, which is
qualitatively equivalent to the discrete one, a fixed number $c$ between
zero and one is chosen, which one calls \textit{the cut}. The cut divides
the interval $[0,1]$ into two parts. Then each agent chooses a value $x_{i}$
between 0 and 1, and the average $x_{m}=\frac{1}{N}\sum_{i}x_{i}$ is
computed. The winning agents are those for which $x_{i}$ lie on the side
opposite to $x_{m}$. That is, the payoff of agent $i$ at time $t$ is 
\begin{equation}
P_{i}(t)=\frac{1}{2}\left( 1-\textnormal{sign}\left\{ \left( x_{m}(t)-c\right)
\left( x_{i}(t)-c\right) \right\} \right)  \label{4.1}
\end{equation}
At each time $t$ the dynamics of agent $i$ is a function of the average
value $x_{m}$ at time $t$ and of a parameter $\alpha _{i}$ that
characterizes his \textit{strategy} 
\begin{equation}
x_{i}(t+1)=f_{i}(x_{m}(t),\alpha _{i})  \label{4.2}
\end{equation}
Here one considers for the function $f_{i}$ either a shifted tent map 
\begin{equation}
f_{i}(x)=2+2x\textnormal{sign}\left( \frac{1}{2}-\left( x+\alpha _{i}\right)
\right) \qquad (\textnormal{mod}1)  \label{4.3}
\end{equation}
or a shifted $p$-ary multiplication 
\begin{equation}
f_{i}(x)=p(x+\alpha _{i})\qquad (\textnormal{mod}1)  \label{4.4}
\end{equation}
$\alpha _{i}$ being a number between zero and one, a different one for each
agent.

At first the strategies, that is the $\alpha _{i}$'s, are randomly chosen.
Then each $r$ time steps, $k$ agents have their strategies modified. The $%
k^{\prime }$ agents with less earnings in that period have new $\alpha $'s
chosen at random and the remaining $k-k^{\prime }$ copy the $\alpha $'s of
the $k-k^{\prime }$ best performers with a small error. This is the
evolution dynamics of this model, the fast dynamics being the one in Eq.(\ref
{4.2}). The variables of the full dynamical systems are $\left( x_{i},\alpha
_{k}\right) $, these variables being coupled by the interplay of fast and
evolution dynamics.

In the discrete minority models originally proposed, each agent has several
strategies at his disposal and at each time step he chooses the strategy
with the best virtual record. The periodic replacement of the worst
strategies by the best ones, used here, is qualitatively equivalent and, in
addition, provides a clear separation between the two types of dynamical
laws that operate in the model. In particular by changing the ratio $\frac{k%
}{r}$ one may explore different time scales for the (\textit{fast}) agent
dynamics and the (\textit{slow}) renewal and copy dynamics. This clear
separation between the two dynamics is important because, as it will be
seen, they play very different roles in the self-organization of the system.

The most interesting feature of the system dynamics is the fact that after a
certain time it approaches a regime where the average value $x_{m}$
oscillates around the value of the cut $c$, even when $c$ is very different
from the random value $0.5$. Fig.5 shows the typical behavior of the system
for $c=0.7$, the map $f_{i}$ being the tent map. The two upper plots show
the approach of $x_{m}$ to the self-organized steady-state through several
steps corresponding to the evolution cycles and the fluctuations of $x_{m}$
around the cut. The number of agents is $N=100$, $k=r=10$ and $k^{^{\prime
}}=3$. The two lower plots in the figure show the distributions of $x_{m}$
and of the fraction of winning agents 
\begin{equation}
P=\frac{1}{N}\sum_{i}P_{i}  \label{4.5}
\end{equation}
The average value and standard deviation of $x_{m}$ are $\overline{x_{m}}%
=0.694$ and $\sigma (x_{m})=0.02$, to be compared with $\overline{x_{m}}=0.5$
and $\sigma (x_{m})=0.288$ that would be obtained for a uniform random
choice of values between 0 and 1 for the agent variables. The fact that $%
x_{m}$ is close to the cut maximizes the percentage of winning agents, which
for the data in Fig.5 is $\overline{P}=0.488$ with $\sigma (P)=0.132$.


The organization of the system's collective variable around the cut $c$ is
easy to understand. Suppose that at a certain time $x_{m}<c$. Then, the
evolution dynamics tends to copy the strategies of the agents that on
average have $x_{i}>c$. This drives the average $x_{m}$ to higher values,
closer to $c$. Conversely if $x_{m}>c$ the effect is just the opposite one.
Hence $x_{m}$ tends to oscillate around $c$. For a minority model with
agents having several strategies at their disposal, the choice of the
strategies with the best virtual record has the same effect. Cavagna\cite
{Cavagna2} pointed out the irrelevance of the memory size (the number of
past time steps that the agents use in their strategies) and stated that the
important issue is that all agents use the same collective information. A
more accurate statement would be that the organization of the collective
variable around $c$ depends only on the evolution dynamics, not on the
details of the (fast) dynamics of the agents.

However, the dynamics of the agents, that is, the nature of their
strategies, has an effect on the type of fluctuations around $c$. The
Lyapunov exponents for the dynamics of $x_{m}$ and for the dynamics of the
agents characterize these fluctuations. The dynamics of $x_{m}$ is 
\begin{equation}
x_{m}(t+1)=\frac{1}{N}\sum_{i}f_{i}\left( x_{m}(t)+\alpha _{i}\right)
\label{4.6}
\end{equation}
the Lyapunov exponent being 
\begin{equation}
\lambda =\lim_{k\rightarrow \infty }\frac{1}{k}\log \left( \frac{1}{N}\left|
\sum_{i}f_{i}^{\prime }(x_{m}(t)+\alpha _{i})\right| \cdots \frac{1}{N}%
\left| \sum_{i}f_{i}^{\prime }(x_{m}(t+k)+\alpha _{i})\right| \right)
\label{4.7}
\end{equation}
For the tent map $\lambda $, assuming, for a large number agents, a uniform
distribution of the $\alpha $'s over the interval, $\frac{1}{N}\left|
\sum_{i}f_{i}^{\prime }\right| $ is of order $\frac{1}{\sqrt{N}}$, hence $%
\lambda $ is negative of order $-\frac{1}{2}\log N$. For the $p$-ary map $%
\lambda =p$.

For the dynamics of the agents the Jacobian matrix is 
\begin{equation}
DT=\left( 
\begin{array}{cccc}
\frac{1}{N}f_{1}^{\prime } & \frac{1}{N}f_{1}^{\prime } & \cdots & \frac{1}{N%
}f_{1}^{\prime } \\ 
\vdots & \vdots &  & \vdots \\ 
\vdots & \vdots &  & \vdots \\ 
\frac{1}{N}f_{N}^{\prime } & \frac{1}{N}f_{N}^{\prime } & \cdots & \frac{1}{N%
}f_{N}^{\prime }
\end{array}
\right)  \label{4.8}
\end{equation}
The eigenvalues of $\left( DT^{k}\right) ^{T}\left( DT^{k}\right) $ are $N-1$
zeros and one equal to 
\begin{equation}
N\left( \frac{1}{N^{2}}\sum_{i}f_{i}^{\prime 2}\right) \left( \frac{1}{N}%
\sum_{i}f_{i}^{\prime }\right) ^{2}\cdots \left( \frac{1}{N}%
\sum_{i}f_{i}^{\prime }\right) ^{2}  \label{4.9}
\end{equation}
Therefore there is only one non-trivial Lyapunov exponent identical to the
Lyapunov exponent of the $x_{m}$ dynamics.

It is the evolution dynamics that organizes the system, driving $x_{m}$
towards the cut. The fast dynamics controls the nature of the fluctuations
around this value. For the tent map, all the Lyapunov exponents being
negative, in the time intervals of duration $r$ between the evolution steps
the dynamics settles down at a fast rate to a fixed point or periodic orbit.
Nevertheless the behavior of the collective variable around its average
value, as shown in Fig.5 is quite irregular. The reason why this is
compatible with the fast contraction associated to negative Lyapunov
exponents is the sensitivity of the attractor to small changes on the agents
strategies, that is, to the variables where the evolution dynamics acts.
Therefore the $\left( x_{i},\alpha _{k}\right) -$dynamical system is a
multi-attractor system. For each fixed set of strategies and initial
conditions the system converges rapidly to a period orbit. However small
changes of the $\alpha -$variables, induced by the evolution process, change
the attractor to which the system converges.


For the $p$-ary maps the existence of one positive Lyapunov exponent changes
the nature of the fluctuations around the mean collective value, meaning
that in this case the system is still a multi-attractor one, but the
attractors are not necessarily periodic. This is shown in Fig.6 where the
same quantities as in Fig.5 are plotted. The fluctuations now partly spoil
the self-organization induced by evolution. The average values and standard
deviations for the data in Fig.6 are $\overline{x_{m}}=0.554$, $\sigma
(x_{m})=0.145$, $\overline{P}=0.378$ and $\sigma (P)=0.223$. Also in this
case, the evolution dynamics controls the collective behavior and acts as a
selector of the attractors of the fast agent dynamics. The difference to the
previous case is that here, rather than periodic orbits, one has
non-periodic attractors.

In contrast to the mechanisms discussed before, where the approach towards $%
0^{+}$ of the positive Lyapunov exponents is the source of the
self-organized collective variables, here it is a multi-attractor evolution
mechanism that determines the emergence of such variables and, on the
contrary, positive Lyapunov exponents may, to some extent, spoil the
self-organization of the system.

\subsection{A market-like game}

In the minority model, as we have seen, the primary interaction between the
agents takes place through an external collective variable, which they
themselves create. In addition there is another dynamical (slower) mechanism
which is the copy of the best strategies by the worst players or,
alternatively the choice by each player of the best virtual strategy among a
number of strategies put at their disposal. This paradigm is very much alike
what happens in a market, where, among other things, investors react to the
stock prices, which they themselves influence through their investments. At
the same time they evolve in time trying to adjust their strategies in order
to maximize their profits. Therefore, once again, we have an interaction
through a collective variable and an evolution mechanism driven by the
desire to maximize a cost function (the profit). From the lessons learned in
the minority model, one is led to expect the collective variables to be
controlled by the evolution mechanism with the agent dynamics providing the
attractor background.

Many factors play a role in a real market. Here no attempt is made to take
into account all the relevant factors, nor to build an accurate model of the
market place. The objective is to isolate some of the mechanisms that
presumably play a role in the market and, by stripping the model from other
(inessential?) complications, to exhibit and understand the \textit{purified}
effect of these factors. In a real market, many factors, endogenous and
exogenous, play a role and one should not expect to find such a clear
cause-effect relationship between dynamical laws and actual behavior.
Nevertheless, as in other branches of science, the splitting apart of the
dynamical components of a phenomena, may improve its understanding \cite
{Roehner}.

We consider a set of investors playing \textit{against }the market, that is
they have some effect on an existing market that is influenced by other
factors (other investors and general economic effects). This assumption
implies that in addition to the impact function of this group of investors
on the market, the rest of the impact is represented by a stochastic
process. Therefore 
\begin{equation}
z_{t+1}=f\left( z_{t},w_{t}\right) +\eta _{t}  \label{4.10}
\end{equation}
represents the change in the log price $\left( z_{t}=\log p_{t}\right) $
with $w_{t}$ being the total investment made by the group of traders and $%
\eta _{t}$ the stochastic process that represents all the other effects.

In addition, no conservation law is assumed for the total amount of \textit{%
stock} $s$ and \textit{cash} $m$ detained by the group of traders. If $p_{t}$
is the price of the traded asset at time $t$, the purpose of the group of
investors is to have an increase, as large as possible, of the total wealth $%
m_{t}+p_{t}\times s_{t}$ at the expense of the rest of the market.

For purposes of comparison with the minority model, here the collective
variable $z$ plays the role of the average value $x_{m}$ and the difference
between the initial wealth and the wealth at time $t$%
\begin{equation}
\Delta _{t}=\sum_{i}\left( m_{t}^{(i)}+p_{t}\times s_{t}^{(i)}\right)
-\sum_{i}\left( m_{0}^{(i)}+p_{0}\times s_{0}^{(i)}\right)  \label{4.11}
\end{equation}
plays the role of the total payoff $P$\medskip

\textbf{The market impact function}\medskip

Let $p$ be the price of some asset, $z=\log (p)$ and $\omega _{t}$ the total
sum of the buying and selling orders (in money units) for the asset. Buying
orders are positive and selling ones negative. An important factor in the
models is the effect of the magnitude of these orders on the price change of
the asset, the so called \textit{market impact function}. Let small orders
have an impact according to the loglinear law\cite{Farmer1} \cite{Bouchaud} 
\begin{equation}
z_{t+1}-z_{t}=\frac{\omega _{t}}{\lambda }+\eta _{t}  \label{4.12}
\end{equation}
The constant $\lambda $, called the \textit{liquidity}, controls the
volatility of the market. It corresponds naturally to a first order
expansion and satisfies the condition 
\begin{equation}
p(p(p_{0},\omega ^{(1)}),\omega ^{(2)})=p(p_{0},\omega ^{(1)}+\omega ^{(2)})
\label{4.13}
\end{equation}
which one expects to be valid for small orders. However, as pointed out by
Zhang\cite{Zhang3} there is experimental evidence that this is not an
accurate representation for large orders. Therefore a slightly different
market impact function will be used. The reasoning used to motivate it has
some relation to Zhang's although the result is somewhat different.

When using Eq.(\ref{4.12}) in a discrete dynamics model we are somehow
neglecting the fact that the market takes different times to fulfill (and to
react to) small and large orders. Therefore this should be taken into
account when reducing the dynamics to a sequence of equal time steps. In
particular the reaction of the market may be parametrized by a change in the 
$\lambda $ coefficient, which being related in first approximation to a
random walk may vary by a factor proportional to $\sqrt{t}$. Taking the time 
$t$ to fill an order to be proportional to its size, one obtains 
\begin{equation}
z_{t+1}-z_{t}=\frac{\omega _{t}}{\lambda _{0}+\lambda _{1}\left| \omega
_{t}\right| ^{1/2}}+\eta _{t}  \label{4.14}
\end{equation}
For small orders one recovers the loglinear approximation and for very large
orders Zhang's square root law.\medskip

\textbf{The agent strategies}\medskip

In first-order, two main types of informations are taken into account by the
investors, namely the difference between price and perceived actual value
(the misprice) 
\begin{equation}
zv_{t}-z_{t}=\log (v_{t})-\log (p_{t})  \label{4.15}
\end{equation}
and the variation in time of the price (the price trend) 
\begin{equation}
z_{t}-z_{t-1}=\log (p_{t})-\log (p_{t-1})  \label{4.16}
\end{equation}
One may also consider differences of prices going further back in time.
However, the qualitative effect on the dynamics is basically the same and,
in line with the main aim of isolating the fundamental constituents of the
process, only these two pieces of informations will be considered. Consider
now a non-decreasing function $f(x)$ such that $f(-\infty )=0$ and $f(\infty
)=1$. Two useful examples are 
\begin{equation}
\begin{array}{lll}
f_{1}(x) & = & \theta (x) \\ 
f_{2}(x) & = & \frac{1}{1+\exp (-\beta x)}
\end{array}
\label{4.17}
\end{equation}
The information about misprice and price trend is coded on a four-component
vector $\gamma $%
\begin{equation}
\gamma _{t}=\left( 
\begin{array}{c}
f(zv_{t}-z_{t})f(z_{t}-z_{t-1}) \\ 
f(zv_{t}-z_{t})\left( 1-f(z_{t}-z_{t-1})\right) \\ 
\left( 1-f(zv_{t}-z_{t})\right) f(z_{t}-z_{t-1}) \\ 
\left( 1-f(zv_{t}-z_{t})\right) \left( 1-f(z_{t}-z_{t-1})\right)
\end{array}
\right)  \label{4.18}
\end{equation}
The strategy of each investor is also a four-component vector $\alpha ^{(i)}$
with entries $-1,$ $0,$ or $1$. $-1$ means to sell, $1$ means to buy and $0$
means to do nothing. Hence, at each time, the investment of agent $i$ is $%
\alpha ^{(i)}\cdot \gamma $ . A fundamental (value-investing strategy) that
buys when the price is smaller than the value and sells otherwise would be $%
\alpha ^{(i)}=\left( 1,1,-1,-1\right) $ and a pure trend-following
(technical trading) strategy would be $\alpha ^{(i)}=\left( 1,-1,1,-1\right) 
$ . In this setting the total number of possible strategies is $3^{4}=81$.
For future reference the strategies will be labelled by a number 
\begin{equation}
n^{(i)}=\sum_{k=0}^{3}3^{k}\left( \alpha _{k}^{(i)}+1\right)  \label{4.19}
\end{equation}
Therefore the fundamental strategy is strategy no. $72$ and the pure
trend-following one is no. $60$.

As compared with a realistic market model, an important ingredient that is
missing is to take into account transaction costs. In actual practice
however excessive transaction costs are avoided by the introduction of
thresholds in the agent strategies. Therefore neglecting both thresholds and
transaction costs has a compensating effect and, qualitatively, the behavior
is expected to be the same.

Another factor that is sometimes considered in market models is the
dependence of the strategies on the values of the prices not only at $t$ and 
$t-1$ but also on a larger set of past times. However, if the lesson that is
learned from minority models also applies here, the size of the memory and
the details of the agents strategies are not very important as far as the
collective properties of the model are concerned. What seemed to be
important there was the copy mechanism in the evolution dynamics and the
value of the Lyapunov exponents to control the fluctuations around (and
away) from the value of the collective variable, the latter being mainly
controlled by the evolution dynamics.

The evolution dynamics that is considered here for the market model is
similar to the one of the minority model. After a number $r$ of time steps, $%
s $ agents copy the strategy of the $s$ best performers and, at the same
time, have some probability to mutate that strategy. This evolution aims at
attaining the goal of improving gains, while at the same time allowing for
some renewal of the strategies. The percentage of each strategy changes in
time and one may find whether some of them become dominating or stable and
when this may occur.

Figs.7 to 10 show the results of some simulations of the model. The
parameters that were kept fixed are $r=50$, $s=10$, $\lambda _{0}=10000$.
The simulations differ by the choice of the initial conditions and the
existence or non-existence of evolution. For Fig.7 an initial condition is
chosen with all traders in the fundamental strategy and evolution is
activated. It is seen that on average the price follows value, although its
fluctuations are amplified. In the last plot of Fig.7 one shows the time
evolution of the strategies distribution coded according to (\ref{4.19}). The
fundamental strategy is seen to be stable, in the sense that it 
becomes dominant, not
being invaded by any other of the strategies that are created by the mutation
process. There are however a few other strategies that, after being created,
survive the selection process. This is true for example for the strategies $%
45=\left( 0,1,-1,-1\right) $, $18=\left( -1,1,-1,-1\right) $, $73=\left(
1,1,-1,0\right) $ and $75=\left( 1,1,0,-1\right) $. These surviving
strategies are however similar to the fundamental one. When there is
dominance of the fundamental strategies, the price increments $dp$ have a
Gaussian distribution. On the other hand the collective objective of
increasing gains $\Delta _{t}$ (Eq.(\ref{4.11})) is achieved, as shown in
the third plot of Fig.7.


For the simulation of Fig.8 the initial condition contains $50\%$ of
fundamental strategies (no. 72) and $50\%$ of trend-following ones (no. 60).
It is seen that the trend following strategies do not survive the selection
process and are eliminated, after a transient period. The statistically
stable situation that is obtained is similar to the one shown in Fig.7.
However the fundamental strategy ceases to be stable if it occurs in the
initial condition in smaller amounts $\left( \leq 40\%\right) $. The
dependence on the initial condition is manifest in Fig.9, where one starts
for a completely random mixture of strategies in the initial condition. In
this case, although the selection mechanism is still favoring at each
evaluation cycle the best performers, the system never organizes itself to
make $\Delta _{t}$ grow.



Finally the simulation in Fig.10 is performed without evolution, with a
fixed $50\%$ of fundamental strategies (no. 72) and $50\%$ of
trend-following ones (no. 60). One sees in this case a large number of
bubbles and crashes in the price evolution and the price increments
distribution has fat tails. The last plot in Fig.10 is an expanded plot of
the bubble around time step $39800$.


To understand the nature of the dynamics that leads to the results of the
simulations is useful to compute the Lyapunov exponents for the log-price $%
\left( z_{t}\right) $ dynamics. The Jacobian for the dynamics 
\begin{equation}
\left( 
\begin{array}{l}
z_{t} \\ 
z_{t-1}
\end{array}
\right) \rightarrow \left( 
\begin{array}{l}
z_{t+1} \\ 
z_{t}
\end{array}
\right)  \label{4.20}
\end{equation}
is 
\begin{equation}
M_{t}=\left( 
\begin{array}{cc}
1+\frac{\partial }{\partial z_{t}}\frac{\sum_{i}\omega ^{(i)}}{\lambda
_{0}+\lambda _{1}\left| \sum_{i}\omega ^{(i)}\right| } & \frac{\partial }{%
\partial z_{t-1}}\frac{\sum_{i}\omega ^{(i)}}{\lambda _{0}+\lambda
_{1}\left| \sum_{i}\omega ^{(i)}\right| } \\ 
1 & 0
\end{array}
\right)  \label{4.21}
\end{equation}
the Lyapunov spectrum being obtained from 
\begin{equation}
\lim_{N\rightarrow \infty }\left| M_{t+N-1}^{T}\cdots M_{t}^{T}M_{t}\cdots
M_{t+N-1}\right| ^{1/2N}  \label{4.22}
\end{equation}
Lyapunov exponents were computed for $f=f_{2}$ (Eq.(\ref{4.17})) for several
values of $\beta $ and a $50-50$ admixture of fundamental and
trend-following strategies. Typically, that is for a very large range of $%
\beta $, one obtains one Lyapunov number equal to zero and the other smaller
but close to one. Therefore, typically, one has two negative Lyapunov
exponents, one of them close to zero.

Although more complex than the minority model, the similarities are very
clear. In the minority model there is an \textit{objective variable} (the
payoff $P$) which drives the evolution and a \textit{collective variable} $%
x_{m}$ to which the agents react in the short run. Here the objective
variable that controls the evolution is $p\times s+m$ and the price is the
collective variable. In both cases the average behavior of the collective
variable is controlled by evolution and the fluctuations by the (fast) agent
dynamics. Whether the objective variable reaches a stable behavior depends
of course on the interaction between the two dynamical laws. Multistability
of the coupled dynamics, rather than the Lyapunov spectrum, seems in both
cases to be the main structure-generating mechanism.

\section{Appendix. Some parameters characterizing the dynamics of
multi-agent systems}

In this appendix one collects the definitions and some properties of a few
parameters which may be used to characterize in a quantitative manner the
self-organization of multi-agent systems. Two different cases are
considered. The first concerns systems where only one invariant measure
controls the dynamics and the second is the case where many different
measures come into play.

\subsection{Ergodic invariants}

Let a dynamical system evolve on the support of some measure $\mu $ which is
left invariant by the dynamics. An \textit{ergodic invariant} is a dynamical
characterization of this measure 
\begin{equation}
I\left( \mu \right) =\lim_{T\rightarrow \infty 
}\frac{1}{T}\sum_{n=1}^{\textnormal{%
T}}\digamma \left( f^{n}x_{0}\right)  \label{A.0}
\end{equation}
for $x_{0}$ $\mu -$almost everywhere.

\subsubsection{Lyapunov and conditional exponents}

Let $f:M\rightarrow M$\ , with $M\subset R^{m}$, $\mu $ a measure invariant
under $f$ and a splitting of $M$ induced by $\Sigma =R^{k}\times R^{m-k}$.
The \textit{conditional exponents}\cite{Pecora} \cite{Vilela1}\ are the
eigenvalues $\xi _{i}^{(k)}$ and $\xi _{i}^{(m-k)}$ of the limits 
\begin{eqnarray}
&&\lim_{n\rightarrow \infty }\left( D_{k}f^{n*}(x)D_{k}f^{n}(x)\right) ^{%
\frac{1}{2n}}  \label{A.1} \\
&&\lim_{n\rightarrow \infty }\left( D_{m-k}f^{n*}(x)D_{k}f^{n}(x)\right) ^{%
\frac{1}{2n}}  \nonumber
\end{eqnarray}
where $D_{k}f^{n}$\ and $D_{m-k}f^{n}$ are the $k\times k$\ and $m-k\times
m-k$ diagonal blocks of the full Jacobian. For $k=m$ , $\xi
_{i}^{(m)}=\lambda _{i}$ are the \textit{Lyapunov exponents}. Both the
Lyapunov and the conditional exponents are ergodic invariants, with
existence $\mu $-almost everywhere guaranteed by the conditions of
Oseledec's multiplicative ergodic theorem, in particular the integrability
condition 
\begin{equation}
\int \mu (dx)\log ^{+}\left\| T(x)\right\| <\infty  \label{A.2}
\end{equation}
$T$ being either the Jacobian or its $k\times k$\ and $m-k\times m-k$
diagonal blocks. The set of regular points is Borel of full measure and 
\begin{equation}
\lim_{n\rightarrow \infty }\frac{1}{n}\log \left\| D_{k}f^{n}(x)u\right\|
=\xi _{i}^{(k)}  \label{A.3}
\end{equation}
\ with $0\neq u\in E_{x}^{i}/E_{x}^{i+1}$\ , $E_{x}^{i}$\ being the subspace
of $R^{k}$\ spanned by eigenstates corresponding to eigenvalues $\leq \exp
(\xi _{i}^{(k)})$.

\subsubsection{Dynamical selforganization}

Self-organization in a system concerns the dynamical relation of the whole
to its parts. The conditional Lyapunov exponents, being quantities that
separate the intrinsic dynamics of each component from the influence of the
other parts in the system, provide a \textit{measure of dynamical
selforganization} $I(S,\Sigma ,\mu )$%
\begin{equation}
I(S,\Sigma ,\mu )=\sum_{k=1}^{N}\left\{ h_{k}(\mu )+h_{m-k}(\mu )-h(\mu
)\right\}  \label{A.4}
\end{equation}
\ the sum being over all relevant partitions $R^{k}\times R^{m-k}$ and$\,$

$h_{k}(\mu )=\sum_{\xi _{i}^{(k)}>0}\xi _{i}^{(k)}$ ; $h_{m-k}(\mu
)=\sum_{\xi _{i}^{(m-k)}>0}\xi _{i}^{(m-k)}$ ; $h(\mu )=\sum_{\lambda
_{i}>0}\lambda _{i}$

\subsubsection{Conditional entropies}

Consider cylindrical partitions adapted to the splitting $R^{k}\times
R^{m-k} $, 
\begin{equation}
\begin{array}{lll}
\eta ^{(k)} & = & \left\{ C_{1}^{(k)},C_{2}^{(k)},\cdots \right\} \\ 
\eta ^{(m-k)} & = & \left\{ C_{1}^{(m-k)},C_{2}^{(m-k)},\cdots \right\}
\end{array}
\label{A.5}
\end{equation}
where $C_{i}^{(k)}$ and $C_{i}^{(m-k)}$ are $k$ and $m-k$ -dimensional
cylinder sets in $R^{m}$.

Let now $\zeta $ be a generator partition for the dynamics $\left\{ f,\mu
\right\} $. The \textit{conditional entropies }associated to the splitting%
\textit{\ }$R^{k}\times R^{m-k}$ are
\begin{equation}
\begin{array}{lll}
\mathfrak{h}^{(k)} & = & \underset{\eta ^{(k)}}{\sup }\textnormal{ }\underset{%
n\rightarrow \infty }{\lim }\frac{1}{n+1}H\left( \zeta \vee f^{-1}\zeta \vee
\cdots \vee f^{-n}\zeta \mid \eta ^{(k)}\right) \\ 
\mathfrak{h}^{(m-k)} & = & \underset{\eta ^{(m-k)}}{\sup }\textnormal{ 
}\underset{%
n\rightarrow \infty }{\lim }\frac{1}{n+1}H\left( \zeta \vee f^{-1}\zeta \vee
\cdots \vee f^{-n}\zeta \mid \eta ^{(m-k)}\right)
\end{array}
\label{A.6}
\end{equation}
$H\left( \chi \mid \eta \right) $ being 
\begin{equation}
H\left( \chi \mid \eta \right) =-{\int }_{M/\eta }\sum_{i}\mu
\left( C_{i}^{(\chi )}\mid \eta \right) \ln \mu \left( C_{i}^{(\chi )}\mid
\eta \right) d\mu  \label{A.7}
\end{equation}
That is, the conditional entropies are the supremum over all cylinder
partitions of the sum of the conditional Kolmogorov-Sinai entropies.

\subsubsection{The structure index}

A structure (in a collective system) is a phenomenon with a characteristic
scale very different from the scale of the component units in the system. A
structure in space is a feature at a length scale larger than the
characteristic size of the components and a structure in time is a
phenomenon with a time scale larger than the cycle time of the individual
components. A (temporal)\textit{\ structure index} may then be defined by

\begin{equation}
S=\frac{1}{N}\sum_{i=1}^{N_{s}}\frac{T_{i}-T}{T}  \label{A.8}
\end{equation}
where $N$ is the total number of components (agents) in the coupled system, $%
N_{s}$ is the number of structures, $T_{i}$ is the characteristic time of
structure $i$ and $T$ is the cycle time of the isolated components (or,
alternatively the characteristic time of the fastest structure). A similar
definition applies for a \textit{spatial structure index}, by replacing
characteristic times by characteristic lengths.

Structures are collective motions of the system. Therefore their
characteristic times are the characteristic times of the separation
dynamics, that is, the inverse of the positive Lyapunov exponents. Hence,
for the temporal structure index, one may write 
\begin{equation}
S=\frac{1}{N}\sum_{i=1}^{N_{+}}\left( \frac{\lambda _{0}}{\lambda _{i}}%
-1\right)  \label{A.9}
\end{equation}
the sum being over the positive Lyapunov exponents $\lambda _{i}$. $\lambda
_{0}$ is the largest Lyapunov exponent of an isolated component or some
other reference value.

The temporal structure index diverges whenever a Lyapunov exponent
approaches zero. Therefore the structure index diverges at the points where
long time correlations develop.

More details on the construction of the ergodic invariants and their
interpretation as relevant properties of multi-agent systems may be found in 
\cite{Vilela3}.

\subsection{Multistability parameters}

If a dynamical system has multiple attractors for the same set of
parameters, then each attractor will have its own invariant measure. However
in this case these measures are not of great practical interest. Instead, a
global (Lebesgue) measure $\mu $ is defined in phase space, with respect to
which the probability to be in the basin of attraction of each one of the
attractors is computed. Several parameters may be used to characterize the
multistable system.

One is the \textit{number }$n_{A}(N)$ \textit{of distinct attractors} as a
function of the number $N$ of degrees of freedom (number of agents) of the
system. Alternatively one may define the \textit{scaling function for the
number of attractors} $g_{n}(N)$ such that 
\begin{equation}
\lim_{N\rightarrow \infty }\frac{n_{A}(N)}{g_{n}(N)}=\textnormal{constant}
\label{A.10}
\end{equation}

The diversity of possible dynamical behaviors when the initial conditions
are chosen at random is characterized by the \textit{attractor entropy} 
\begin{equation}
S(N)=\sum_{i}\mu (b_{i})\log \mu (b_{i})  \label{A.11}
\end{equation}
$b_{i}$ being the \textit{basin of attraction} corresponding to the $i$
attractor. As in (\ref{A.10}) a scaling function $g_{A}(N)$ may be defined
for the entropy.

When a multistable system is perturbed, by noise or by fluctuations in the
parameters, migration between attractors takes place which, in addition to
the intensity of the perturbation, is strongly influenced by the stability
of the attractors and by the nature of the boundaries of the basins of
attraction. Given a metric in phase space, the stability of the attractors
may be characterized by its \textit{average} \textit{strength }$\overline{s}$
defined as the average of the minimum distances $d_{\min }(i)$ between the
attractors and the boundary of their basins of attraction, scaled by average
size of a basin of attraction 
\begin{equation}
\overline{s}=\frac{1}{n_{A}(N)^{1-\frac{1}{d}}}\sum_{i}d_{\min }(i)
\label{A.12}
\end{equation}
$d$ being the geometrical dimension of the phase space.

Another important factor controlling attractor migration is the \textit{%
Hausdorff dimension of the basin boundaries}. If this dimension is high (in
some cases it may approach $d$) the noise-perturbed system may spend most of
the time in such a riddled boundary, without ever setting in any particular
attractor. This situation leads to a high degree of unpredicability, even
higher than the usual chaotic (positive Lyapunov exponent) regime.

A easier to measure characterization of the effect of attractor strength and
basin boundary structure on the migration dynamics is the \textit{mean
first-passage time} $\overline{\tau }(\varepsilon )$ between attractors as a
function of the noise intensity $\varepsilon .$

\end{document}